\documentclass[preprint,showpacs,preprintnumbers,amsmath,amssymb]{revtex4}

\usepackage{graphicx}
\usepackage{dcolumn}
\usepackage{bm}

\begin{document}

\title{Matrix exponential solution of the Landau-Zener problem}
\author{Alberto G. Rojo}
\email{rojo@oakland.edu}
\affiliation{%
Department of Physics, Oakland University, Rochester, MI 48309.}

  \begin{abstract}
We present a derivation of the Landau-Zener solution through the
explicit evaluation of the time ordered propagator.  The result is
exact and does not involve the solution of the differential equation
for the spin amplitudes.
  \end{abstract}
  \maketitle

  In 1932, C. Zener presented an exact solution
  for the non-adiabatic transition probability of a two level problem\cite{landauzener}.
Landau solved it simultaneously (with an error factor of
$2\pi$)\cite{landauzener2} and the problem--one of the few cases
where a non-adiabatic transition can be computed exactly--became a
classic and is now called the Landau-Zener (LZ) model.

Although the time dependent problem for two level systems is simple
at first sight, the exact solutions are known for a handful of
cases. One is the LZ model considered in this paper, where the level
spacing $\epsilon_1-\epsilon_2$  varies linearly with time and level
coupling between them, $\epsilon_{12}$, is constant. On the other
hand there is a class of models for which the level spacing is
constant and the coupling is $\epsilon_{12}=g(t)$ with $g$ an
integrable function. Surprisingly the only cases of a  smooth and
even $g$  for which the solution is known is $g(t)=\lambda /\cosh
(t/\beta)$ and some specific time-dependent energies and couplings
.\cite{bambini,berman}

Zener solves Schr\"odinger's equation for all times in terms of
Weber's functions and then computes the exact transition
probability. This is the basis of most approaches, with the
exception (to our knowledge) of the recent paper by C.
Wittig\cite{wittig}, who derives the Landau-Zener formula using
contour integration, bypassing the solution of the differential
equation.

In this paper we present a different derivation of the transition
probability by direct evaluation of the time-ordered matrix
propagator. The solution is exact, does not involve any
approximation, and does not make reference to the differential
equation. Instead, we solve the time ordered propagator with a
couple of simple manipulations and obtain the transition probability
(and not the full time dependence of the problem). Since the
Landau-Zener has innumerable applications to quantum problems and is
the subject of current approaches (theoretical and experimental) in
solid-state artificial atoms\cite{berns}, quantum
information\cite{LaHaye}, quantum dissipation\cite{Nalbach} (to name
just a a few cases), we believe that this new derivation of an old
result can be useful and hopefully motivate new avenues for the
problem. In addition, it is a pleasure to solve an old problem from
a new perspective.

The following is the Schr\"odinger equation of Landau-Zener
problem\cite{landauzener}.

\begin{equation}
i\hbar {d\over dt}\left(\begin{array}{c} a(t)\\b(t)\end{array}
\right)= \left(\begin{array}{cc} \alpha t& \Gamma\\
\Gamma & - \alpha t
\end{array} \right) \left(\begin{array}{c} a(t)\\b(t)\end{array}
\right).
\end{equation}

It is interesting that, for
\begin{equation}a(-\infty)=1, \;\;\;\; b(-\infty)=0,\end{equation} the exact solution gives

$$|a(\infty)|^2=e^{-\pi\Gamma^2/\hbar \alpha}
$$

The fact that the solution is an exponential of the relevant
dimension-less parameter suggests (to us) that there should be a
simple derivation.

First transform the hamiltonian to
$$
\overline{H}=UHU^*$$and the spinor to

$$
\left(\begin{array}{c} \overline{a}(t)\\\overline{b}(t)\end{array}
\right)= U\left(\begin{array}{c} {a}(t)\\{b}(t)\end{array} \right)
$$
with \begin{equation} U= \left(\begin{array}{cc} e^{i\alpha
t^2/2\hbar}&0\\0& e^{-{i\alpha t^2/ 2\hbar}} \end{array} \right).
\end{equation}

We obtain

\begin{eqnarray}
i\hbar {d\over dt}\left(\begin{array}{c}
\overline{a}\\\overline{b}\end{array} \right)&=&
\left(\begin{array}{cc} { } 0& \Gamma e^{-i\alpha
t^2/\hbar}\\
\Gamma e^{i\alpha t^2/\hbar}& 0
\end{array} \right) \left(\begin{array}{c} \overline{a}\\\overline{b}\end{array}
\right)\nonumber
\\
&\equiv& \overline{H} (t)\left(\begin{array}{c}
\overline{a}\\\overline{b}\end{array} \right). \label{HH}
\end{eqnarray}

The formal solution of (\ref{HH}) is

\begin{eqnarray}
\left(\begin{array}{c}
\overline{a}(\infty)\\\overline{b}(\infty)\end{array} \right)&=& T
e^{-{i\over \hbar}\int_{-\infty}^{\infty} dt \overline{H}(t)}
\left(\begin{array}{c}
\overline{a}(-\infty)\\\overline{b}(-\infty)\end{array} \right),
\end{eqnarray}

with
 \begin{equation}
 T
e^{-{i\over \hbar}\int_{-\infty}^{\infty} dt \overline{H}(t)} =1-
{i\over \hbar} \int_{-\infty}^{\infty} dt \overline{H}(t) - {1\over
\hbar^2} \int_{-\infty}^{\infty} dt_1
\overline{H}(t_1)\int_{-\infty}^{t_1} dt_2 \overline{H}(t_2)+\cdots
  \end{equation}
the time ordered exponential. The time ordering is crucial since the
two dimensional matrices $\overline{H}(t)$ do not commute at
different times.

Note that the even powers of $\overline{H}$ have diagonal terms only
whereas the odd powers have only off-diagonal terms. For example

\begin{eqnarray}
\overline{H}(t_1) \overline{H}(t_2) \overline{H}(t_3)&=&
\left(\begin{array}{cc} 0&\Gamma^3e^{-i\alpha (t_1^2-t_2^2+t_3^2)/\hbar}\\
\Gamma^3e^{i\alpha (t_1^2-t_2^2+t_3^2) /\hbar} &0\end{array}
\right),
 \nonumber
\\
\overline{H}(t_1) \overline{H}(t_2) \overline{H}(t_3)
\overline{H}(t_4)&=&
\left(\begin{array}{cc} \Gamma^4e^{-i\alpha (t_1^2-t_2^2+t_3^2-t_4^2)/\hbar}&0\\
0&\Gamma^4e^{i\alpha (t_1^2-t_2^2+t_3^2-t_4^2) /\hbar} \end{array}
\right). \nonumber
\end{eqnarray}

This means that, if our initial condition is
$\overline{a}(-\infty)=1$, $\overline{b}(-\infty)=0$, the solution
for $\overline{a}$ requires only the diagonal term and will be given
by

\begin{eqnarray}
\overline{a}(\infty)&=&1-\left(\Gamma\over \hbar
\right)^2\int_{-\infty}^{\infty}dt_1 e^{-i\alpha t_1^2/\hbar}
\int_{-\infty}^{t_1}dt_2 e^{i\alpha t_2^2/\hbar}\nonumber
\\
&&+\left(\Gamma\over \hbar \right)^4\int_{-\infty}^{\infty}dt_1
e^{-i\alpha t_1^2/\hbar}
 \int_{-\infty}^{t_1}dt_2
 e^{i\alpha t_2^2/\hbar}
\int_{-\infty}^{t_2}dt_3 e^{-i\alpha t_3^2/\hbar}
 \int_{-\infty}^{t_3}dt_4 e^{i\alpha
t_4^2/\hbar} +\cdots \;.\nonumber
\end{eqnarray}
or, in terms of the  quantity \begin{equation}\gamma ={
\Gamma^2\over \hbar \alpha},\end{equation}

\begin{eqnarray}
\overline{a}(\infty)&=&1-\gamma\int_{-\infty}^{\infty}dt_1 e^{-i
t_1^2} \int_{-\infty}^{t_1}dt_2 e^{i t_2^2}\nonumber
\\
&&+\gamma^2\int_{-\infty}^{\infty}dt_1 e^{-i t_1^2}
 \int_{-\infty}^{t_1}dt_2
 e^{i t_2^2}
\int_{-\infty}^{t_2}dt_3 e^{-it_3^2}
 \int_{-\infty}^{t_3}dt_4 e^{i
t_4^2} +\cdots \nonumber \\&\equiv& \sum_{n=0}^\infty
(-\gamma)^nT_{2n}. \label{ainfty}
\end{eqnarray}

Since the time ordered integrals $T_{2n}$ involve alternating
exponentials $e^{it^2}$ and $e^{-it^2}$ we cannot use simple
permutation symmetry of the indices to evaluate them. This in
contrast with a time ordered integral $I_n$ of $n$ identical
functions $f(t)$, where we have:
\begin{equation}
I_n=\int_{-\infty}^{\infty}dt_1 f(t_1)
 \int_{-\infty}^{t_1}dt_2
 f(t_2)
 \cdots
\int_{-\infty}^{t_{n-1}}dt_n f(t_n) ={1\over n!}\left[
\int_{-\infty}^{\infty}dt f(t) \right]^n .
\end{equation}

Time ordered integrals can be written using the the step function
$\Theta$ defined as

\begin{equation}
\Theta (t)
=\left\{\begin{array}{c}1\;\;\;\rm{for}\;\;t>0 \\
0\;\;\;\rm{for}\;\;t<0 \end{array}\right. .\end{equation}

Using $\Theta$, the above integral of identical functions can be
written as
\begin{equation}
I_n=\int_{-\infty}^{\infty}\left(\prod_{i=1}^n dt_i\right) f(t_1)
 \Theta(t_1-t_2)
 f(t_2)\Theta(t_2-t_3) \cdots
\Theta(t_{n-1}-t_n)f(t_n) \equiv {\left(I_1\right)^n\over n!}.
\label{InThet}
\end{equation}

In order to evaluate the time ordered integrals $T_{2n}$ of
alternating exponentials we use the functions $\Theta$ in its
integral representation:

\begin{equation}
\Theta (t)={1\over 2\pi i}\int_{-\infty}^{\infty}d\omega {e^{i\omega
t}\over \omega + i\epsilon} \label{ThetRep}.
\end{equation}

Using $\Theta$, the expression for $T_{2n}$ becomes:
\begin{eqnarray}
T_{2n}&=&\int_{-\infty}^{\infty}\left(\prod_{i=1}^{2n} dt_i\right)
e^{it_1^2}
 \Theta(t_1-t_2)
 e^{-it_2^2}\Theta(t_2-t_3) \cdots
e^{it_{2n-1}^2}\Theta(t_{2n-1}-t_{2n})e^{-it_{2n}^2}\nonumber \\
&=&{1\over (2\pi i)^{2n-1} } \left(\sqrt{\pi}\right)^{2n}
\int_{-\infty}^{\infty}\left(\prod_{i=1}^{2n-1} d\omega_i\right)
{1\over \overline{\omega}_1}{e^{i\omega_2(\omega_1-\omega_3)}\over {
\overline{\omega}_2} } {1\over \overline{\omega}_3}
{e^{i\omega_4(\omega_3-\omega_5)}\over  \overline{\omega}_4 }\cdots
{1\over \overline{\omega}_{2n-1}}, \nonumber
\\
\label{T2n}
\end{eqnarray}
where, to abbreviate notation we have introduced
\begin{equation}
\overline{\omega}_k=\omega_k+i\epsilon.
\end{equation}
Also, in going from the first to the second line of (\ref{T2n}) we
completed squares and used the gaussian integral
\begin{equation}
\int_{-\infty}^{\infty}dt \,e^{\pm it^2}= \sqrt{\pi\over \pm i}.
\end{equation}
Since we have an equal number of $+i$'s and $-i$'s in the integrals
we get the positive factor $(\sqrt{\pi})^{2n}$.  Also notice that,
since we have alternating signs in the exponents, the terms
quadratic  in $\omega_k$ vanish. This is the feature that makes the
integral $T_{2n}$ easily solvable.

 Next we perform the
integrals in (\ref{T2n}) over the $n-1$ frequencies $\omega_k$ with
even index $k$ and use the fact that [see Eq.(\ref{ThetRep})]
\begin{equation}
\int_{-\infty}^{\infty}d\omega_k{e^{i\omega_k(\omega_{k-1}-\omega_{k+1})}\over
{ {\omega}_k+i\epsilon} }=2\pi i\Theta(\omega_{k-1}-\omega_{k+1}
)
\label{ThetOm}.
\end{equation}
If we replace (\ref{ThetOm}) in (\ref{T2n}) we are left with $n$
integrals. After relabeling the odd indices to consecutive indices
we get:
\begin{eqnarray}
T_{2n}&=&{1\over (2\pi i)^{n} } \left(\sqrt{\pi}\right)^{2n}
\int_{-\infty}^{\infty}\left(\prod_{k=1}^{n} d\omega_k\right)
{1\over \overline{\omega}_1}\Theta(\omega_1-\omega_2){1\over {
\overline{\omega}_2} } \Theta(\omega_2-\omega_3) \cdots
\Theta(\omega_{n-1}-\omega_n) {1\over \overline{\omega}_{n}}.
\nonumber
\\
\label{T2nB}
\end{eqnarray}
But this is precisely the structure of Eq. (\ref{InThet}): in this
case we have a ``frequency ordered" integral of identical functions
$1/(\omega_k +i\epsilon)$. The result is simply:

\begin{eqnarray}
T_{2n}&=&{\pi}^{n} {1\over n!}\left({1\over 2\pi
i}\int_{-\infty}^{\infty}d\omega {1\over \omega +i\epsilon}\right)^n
\nonumber \\
&\equiv&  {1\over n!}\left({\pi \over 2}\right)^n.
 \label{T2nC}
\end{eqnarray}

Replacing in (\ref{ainfty}) we obtain
\begin{eqnarray}
\overline{a}(\infty) &=&\sum_{n=0}^{\infty}{1\over
n!}\left(-{\pi\gamma\over 2}\right)^n=e^{-\pi\gamma/2}\nonumber \\
&\equiv & e^{-\pi\Gamma^2/2\hbar \alpha},
\end{eqnarray}
and
\begin{equation}
|a(\infty)|^2=e^{-\pi\Gamma^2/\hbar \alpha},
\end{equation}
which is the exact Landau-Zener result.

We thank Anthony Bloch for useful discussions.

\end{document}